\documentstyle[mprocl]{article}

\input{psfig}
\def\be{\begin{equation}}
\def\ee{\end{equation}}
\def\bea{\begin{eqnarray}}
\def\eea{\end{eqnarray}}

\begin{document}

\title{LIMITING SOLUTIONS OF SEQUENCES OF GLOBALLY REGULAR
AND BLACK HOLE SOLUTIONS IN SU(N)-EYMD THEORIES}

\author{A. SOOD, B. KLEIHAUS, J. KUNZ}

\address{Fachbereich Physik, Universit\"at Oldenburg, Postfach 2503\\
D-26111 Oldenburg, Germany}

\maketitle\abstracts{
Static spherically symmetric solutions in SU(N)-EYM and EYMD theories
are classified by the node numbers of their non-trivial gauge field
functions. With increasing node numbers, the solutions form
sequences, tending to limiting solutions.
The limiting solutions are based on subalgebras of $su(N)$,
consisting of a neutral non-abelian part and a charged abelian part,
belonging to the Cartan subalgebra.}

We consider the $SU(N)$ Einstein-Yang-Mills action
\begin{equation}
S=S_G+S_M=\int L_G \sqrt{-g} d^4x + \int L_M \sqrt{-g} d^4x
\ \label{action}  \end{equation}
with
\begin{equation}
L_G=\frac{1}{16\pi G}R \ , \ \ \
L_M=-\frac{1}{2} {\rm Tr} (F_{\mu\nu} F^{\mu\nu})
\ , \label{lagm} \end{equation}
and with field strength tensor
$F_{\mu\nu}= \partial_\mu A_\nu - \partial_\nu A_\mu
            - i e [A_\mu,A_\nu]$,
gauge field
$A_\mu = \frac{1}{2} \lambda^a A_\mu^a$
and gauge coupling constant $e$.

We employ Schwarz\-schild-like coordinates and adopt
the spherically symmetric metric
\begin{equation}
ds^2=g_{\mu\nu}dx^\mu dx^\nu=
  -{\cal A}^2{\cal N} dt^2 + {\cal N}^{-1} dr^2 
  + r^2 (d\theta^2 + \sin^2\theta d\phi^2)
\ , \label{metric} \end{equation}
with the metric functions ${\cal A}(r)$ and 
${\cal N}(r)=1-\frac{2m(r)}{r}$.

The static spherically symmetric ans\"atze
for the gauge field $A_{\mu}$ of $SU(N)$ EYM theory
are based on the $su(2)$ subalgebras of $su(N)$.
Considering only the embedding
of the $N$-dimensional representation of $su(2)$,
the gauge field ansatz is \cite{kuenzle}
\begin{equation} \label{ansa}
 A_{\mu}^{(N)}  dx^\mu   =  \frac{1}{2e} \left(\!\! 
\begin{array}{ccccc}
(N\!-\!1)\cos\theta d\phi & \omega_1 \Theta & 0 & \ldots & 0 \\
\omega_1 \bar \Theta & (N\!-\!3)\cos\theta d\phi & \omega_2 \Theta & 
\ldots & 0 \\
\vdots & & \ddots & & \vdots \\
0 & \ldots & 0 & \omega_{N\!-\!1} \bar \Theta & (1\!-\!N)\cos\theta d\phi
\end{array}\!\!  \right)
\label{amu} \end{equation}   
with $\Theta = i d \theta + \sin \theta d \phi$,
and $A_0=A_r=0$.
The ansatz contains $N-1$ matter field functions
$\omega_j(r)$.

We introduce the dimensionless coordinate 
$x=er/{\sqrt{4\pi G}}$,
the dimensionless mass function
$\mu=em/{\sqrt{4\pi G}}$,
and the scaled matter field functions
$ u_j = {\omega_j}/{\sqrt{\gamma_j}}$ with $\gamma_j = {j (N - j) } $.
Considering the boundary conditions,
globally regular solutions must satisfy at the origin 
$\mu(0)=0$ and $u_j(0)=\pm 1 $.
Black hole solutions with a regular horizon 
with radius $x_{\rm H}$ must satisfy at the horizon
${\cal N}(x_{\rm H})=0$
and
\begin{equation}
 \left. {\cal N}^{'}u_j^{'} + \frac{1}{2x^2}   
(\gamma_{j+1} u^2_{j+1} - 2 \gamma_{j} u^2_{j} 
+\gamma_{j-1} u^2_{j-1} +2 ) u_j \right|_{x_{\rm H}} =0
\ . \label{bc5} \end{equation}
For asymptotically flat solutions the metric functions ${\cal A}$
and $\mu$ must approach constants at infinity.
We choose ${\cal A}(\infty)=1 $,
$\mu(\infty)$ is the mass of the solutions.
In neutral solutions
the gauge field functions $u_j$ satisfy
$u_j(\infty)=\pm 1$,
i.e.~the gauge field approaches a vacuum configuration.

All $N-1$ gauge field functions are non-trivial for
neutral globally regular and black hole solutions. 
The solutions are labelled by the node numbers
$n_j$ of the functions $u_j$ \cite{kks}.
When the node numbers of one or more gauge field functions tend to infinity,
the solutions approach limiting solutions, which carry a magnetic charge $P$.
Considering black hole solutions with $x_{\rm H} > P$
(or the exterior part of the solutions with $x_{\rm H} < P$),
we observe that 
in these limiting solutions the corresponding gauge field functions
are identically zero \cite{kks}.

When one or more of the $N-1$ gauge field functions 
are identically zero,
magnetically charged solutions are obtained.
To classify the charged black hole solutions
obtained within the ansatz (\ref{amu}),
let us first assume that precisely one gauge field function 
is identically zero, $\omega_{j_{1}} \equiv 0$.
The ansatz then reduces to
$A_\mu^{(N)} dx^\mu =  $
\begin{equation}
\!\left(\! 
\begin{array}{cc}
{\rm \fbox{$ A_\mu^{(j_1)} dx^\mu  $}} &                      \\
                     &  \!{\rm \fbox{$A_\mu^{(N-j_1)} dx^\mu $}}\\
\end{array}
\!\right)
\!+\frac{\cos{\theta}d\phi}{2e}\!\left(\! 
\begin{array}{cc}
{\rm \fbox{$(N\!-\!j_1){\bf 1}_{(j_1)} $}}&     \\
      &\!{\rm \fbox{$ -j_1 {\bf 1}_{(N\!-\!j_1)}$}}   \\            
\end{array}
\!\right)
\ \label{amu1} \end{equation}
where $A_\mu^{(\bar N)}$ denotes the non-abelian
spherically symmetric ansatz for the $su(\bar N)$
subalgebra of $su(N)$
(based on the $\bar N$-dimensional embedding of $su(2)$),
and the second term represents the abelian ansatz for the element 
${h}$ of the Cartan subalgebra of $su(N)$.
The full system of equations thus consists of two non-abelian
and one abelian subsystems, coupled only via the
metric functions.
The non-abelian parts of the solutions
corresponding to $su(\bar {N})$ subalgebras are neutral
with
$\bar u_i(\infty) = \pm 1$.
The charge of the solutions is carried by the abelian subsystem
\cite{yasskin},
\begin{equation}
P^2 = \frac{1}{2}{\rm Tr}\  {h}^2
\ . \label{Pa} \end{equation}
By expanding the element ${h}$ in terms of the
basis $\{\lambda_{n^2-1} \  | \  n=2, \ldots ,N \}$, 
the charge can also be directly read off
the expansion coefficients (see Table~1).

By applying these considerations again to the subalgebras
$su(\bar N)$ of eq.~(\ref{amu1}), we obtain
the general case for $SU(N)$ EYM theory,
where $M$ gauge field functions are identically zero,
$\omega_{j_m} = 0$, $j_m \in \left\{ j_1, j_2, \ldots , j_M\right\}$, 
$j_1 < j_2 <  \cdots < j_M$.
The charge is again carried by the abelian subsystem,
belonging to the Cartan subalgebra, and given by
\begin{equation}
P^2 = \frac{1}{2} {\sum_{m=1}^M} (N-j_m)(N-j_{m-1})(j_m-j_{m-1})
\ \ \ (j_0=0)
\ . \label{su2n} \end{equation}
In the special case where
all gauge field functions are identically zero,
an embedded RN solution is obtained with charge
$P^2 =  (N-1)N(N+1)/6$.
RN black hole solutions exist only for horizon radius
$x_{\rm H}\ge P$, and
the extremal RN solution has $x_{\rm H}=P$.
The same is true for charged non-abelian black holes.
For extremal non-abelian black hole solutions ${\cal N}^{'}=0$,
leading to $\bar u_i(x_{\rm H})= \pm 1 $.
There are no such globally regular charged solutions.

We demonstrate the above general considerations
for $SU(5)$ EYM theory.
Presenting all inequivalent cases (within the ansatz (\ref{amu})),
the classification of the charged $SU(5)$
EYM black hole solutions is given in Table~1.

 \newcommand{\rb}[1]{\raisebox{1.5ex}[-1.5ex]{#1}}
\begin{table}[hp]
\begin{center}
\begin{tabular}{|c|cccc|c|l|p{0.75cm}p{0.75cm}p{0.75cm}p{0.75cm}|} \hline
 &  & & & &  & \multicolumn{1}{ c|}{non-abelian} &
  \multicolumn{4}{ c|}{Cartan subalgebra} \\
\cline{8-11}
\multicolumn{1}{|c|} { \rb{\#} }&
\multicolumn{4}{ c|} { \rb{$u_j,\ j=1\!-\!4$} }&
\multicolumn{1}{ c|} { \rb{$P^2$} }&
\hspace{1.mm} subalgebra &
  $\lambda_3$ & $\lambda_8$ & $\lambda_{15}$ & $\lambda_{24}$ \\
 \hline
0  & 0     & 0     & 0     & 0      &
20 &  & 
$P_3$   &  $P_8$    & $P_{15}$  &  $P_{24}$     \\
 \hline
1a & $u_1$ & 0     & 0     & 0     & 
19 & $su(2)$ & 
0   &  $P_8$    & $P_{15}$  &  $P_{24}$   \\
 \hline
2a & $u_1$ & $u_2$ & 0     & 0     & 
16 & $su(3)$ & 
0   &  0    & $P_{15}$  &  $P_{24}$  \\
2b & $u_1$ & 0     & $u_3$ & 0     & 
18 & $su(2)\oplus su(2)$ &
0   & $\frac{4}{3}P_8$    & $\frac{2}{3}P_{15}$  &  $P_{24}$ \\
 \hline
3a & $u_1$ & $u_2$ & $u_3$ & 0     & 
10 & $su(4)$  &
0   & 0    & 0  &  $P_{24}$ \\
3b & $u_1$ & $u_2$ & 0     & $u_4$ & 
15 & $su(3)\oplus su(2)$ &
0   & 0    & $\frac{5}{4}P_{15}$  &  $\frac{3}{4}P_{24}$\\
\hline
\end{tabular}
\end{center} 
{\bf Table 1}\\
Charged black hole solutions of $SU(5)$
EYM theory ($P_{n^2-1}=\sqrt{\frac{n(n-1)}{2}}$).
\end{table}

Like the neutral solutions of $SU(N)$ EYM theory, 
the charged solutions are labelled
by the node numbers $n_i$ of their (non-vanishing) gauge field functions
$\bar u_i$.
When the node numbers of one or more gauge field functions
tend to infinity, the solutions approach
limiting solutions with higher charge.
In Fig.~1 we present the lowest odd solutions
of case 3a for the sequence $(n,0,0)$
and extremal horizon $x_{\rm H}=\sqrt{10}$,
also in the black hole interior ($x<x_{\rm H}$).

\begin{figure}[hp]
\rule{5cm}{0.2mm}\hfill\rule{5cm}{0.2mm}
\psfig{figure=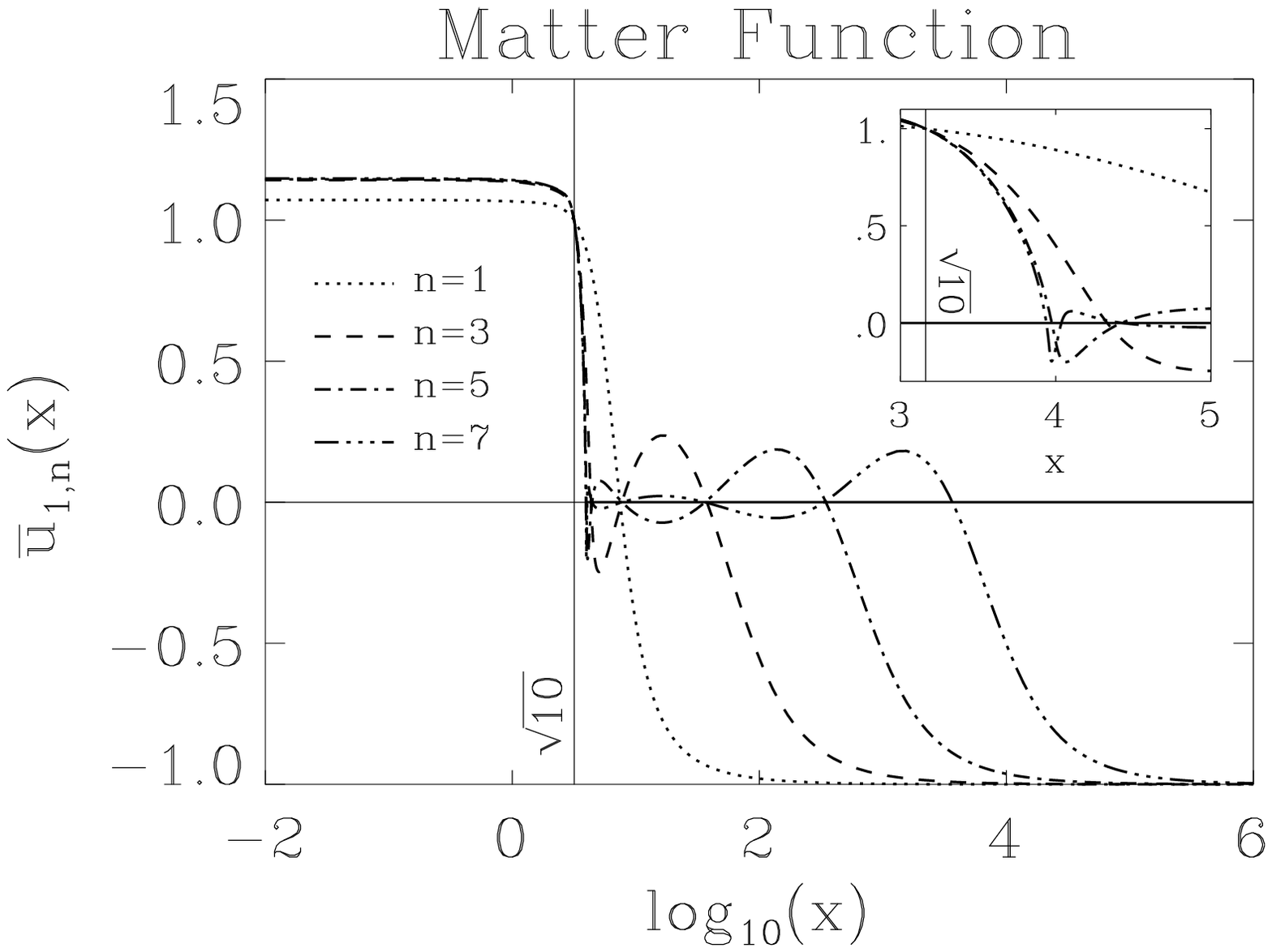,height=1.4in}
\psfig{figure=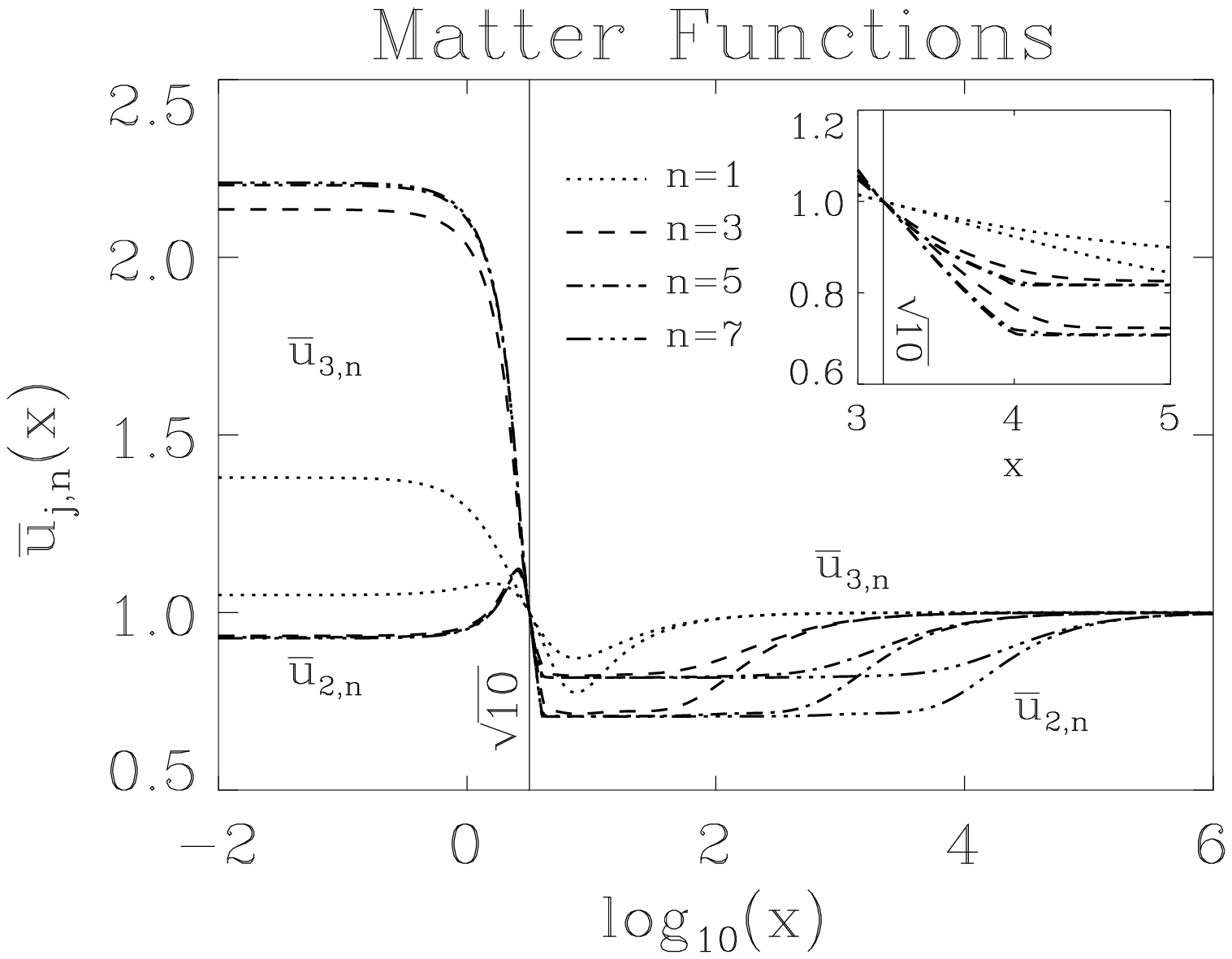,height=1.4in}
\rule{5cm}{0.2mm}\hfill\rule{5cm}{0.2mm}
\caption{
The matter functions 
for case 3a of Table~1 }
\label{fig:fig1}
\end{figure}

The above classification remains valid for SU(N) EYMD theory.
However, the charged non-abelian EYMD black hole solutions exist for
arbitrary horizon radius $x_{\rm H}>0$.

\end{document}